# Nonreciprocal Optical Routing in Multi-port Magneto-Optical Devices on Silicon

XIAOYI SONG[1,2], WEI YAN[3,*], DI WU[1,2], YUCONG YANG[1,2], ZIXUAN WEI[1,2], ZIJIAN ZHANG[1,2], TIANCHI ZHANG[1,2], JUNXIAN WANG[1,2], JUN QIN[1,2], AND LEI BI[1,2,*]

[1]*National Engineering Research Center of Electromagnetic Radiation Control Materials, University of Electronic Science and Technology of China, Chengdu, 611731*

[2]*Key Laboratory of Multi-Spectral Absorbing Materials and Structures of Ministry of Education, University of Electronic Science and Technology of China, Chengdu, 611731*

[3]*School of Optoelectronic Science and Engineering & Collaborative Innovation Center of Suzhou Nano Science and Technology, Soochow University, Suzhou 215006, China*

*\*weiyan085@suda.edu.cn, bilei@uestc.edu.cn*

**Abstract:** Nonreciprocal optical devices are key components in photonic integrated circuits for light reflection blocking and routing. Most reported silicon integrated nonreciprocal optical devices to date were unit devices. To allow complex signal routing between multi-ports in photonic networks, multi-port magneto-optical (MO) nonreciprocal photonic devices are desired. In this study, we report experimental demonstration of a silicon integrated 5×5 multiport nonreciprocal photonic device based on magneto-optical waveguides. By introducing different nonreciprocal phase shift effect to planar photonic waveguides, the device focuses light to different ports for both forward and backward propagation. The device shows designable nonreciprocal transmission between 5×5 ports, achieving 16 dB isolation ratio and -18 dB crosstalk. © 2023 Chinese Laser Press

## 1. Introduction

Driven by the rapid development of data communication and optical communication, multi-port photonic devices such as arrayed waveguide gratings, optical switch matrices and photonic neural networks have been developed [1-5]. As a key component in photonic systems, nonreciprocal photonic devices introduce key functionalities to block the reflected light, or re-route backward transmission signal to a different channel compared to the input [6, 7]. However, most reported silicon integrated nonreciprocal photonic devices to date were unit devices. Typical device structures include Mach-Zehnder interferometers (MZIs) [7-10], multi-mode interferometers (MMIs) [11, 12], and micro-ring resonators [13-15] with a maximum of only four ports, limiting the device functionality to only optical isolation and circulation.

Multi-port silicon integrated nonreciprocal optical devices allow different scattering matrix between forward and backward propagation, which may add another degree of freedom for optical data transmission. For bidirectional optical communication or sensing between multiports, nonreciprocal optical routing is usually achieved by adding bulk optical circulators in each port, which leads to bulky and complex systems [1, 3-5, 10-19]. Therefore, achieving nonreciprocal optical transmission between multi-ports on-chip can greatly simplify the system and introduce design flexibility. So far, the only reported multiport magneto-optical device on silicon was based on cascaded magneto-optical (MO) micro-ring resonators [13], demonstrating optical circulation functionality between six-ports. However, the device showed limited operation bandwidth and similar functionality as an optical circulator. Multi-port nonreciprocal optical devices with larger bandwidths and alternative functionalities are yet to be explored.

In this paper, we report the development of a multi-port magneto-optical router on silicon (Si). we experimentally demonstrated silicon integrated multi-port nonreciprocal optical transmission in a nonreciprocal planar optical focusing structure based on the nonreciprocal phase shift (NRPS) effect [20-25]. On-demand nonreciprocal scattering matrices between forward and backward propagation light are achieved by deposition of magneto-optical thin films on silicon waveguide arrays, introducing nonreciprocal phase profile between forward and backward propagation light. We demonstrated functionalities such as optical circulation and nonreciprocal optical routing between different ports. Experimentally, we demonstrate optical isolation ratio up to 16 dB and a minimum crosstalk of -18 dB at 1551 nm wavelength in a 5×5 multiport nonreciprocal photonic device.

## 2. Operation Principle

Fig. 1(a) shows the sketch of the multi-port nonreciprocal magneto-optical device on 220 nm silicon on insulator (SOI) platforms. The device consists of several input and output ports, two slab Si waveguides connecting to the input and output ports, a phase element array based on Si waveguides providing reciprocal phase shift (RPS), and a MO/Si waveguide array providing nonreciprocal phase shift. Transverse magnetic (TM) polarized light enters the phase element array after passing through the input Si slab waveguide. It then propagates through the phase element array and the MO/Si waveguide array. After accumulating phase shifts consisting of RPS and NRPS in each waveguide in the array, the input light forms a phase gradient at the output of the waveguide array, so that it can be refocused into the corresponding output port [22, 24]. However, when light is reversely incident at the output port, the NRPS accumulated in the MO/Si waveguide array is different from that of the forward direction, so that the backward transmission light can be focused to different ports from the original input port. This mechanism makes it possible to implement on-chip multi-port nonreciprocal routing.

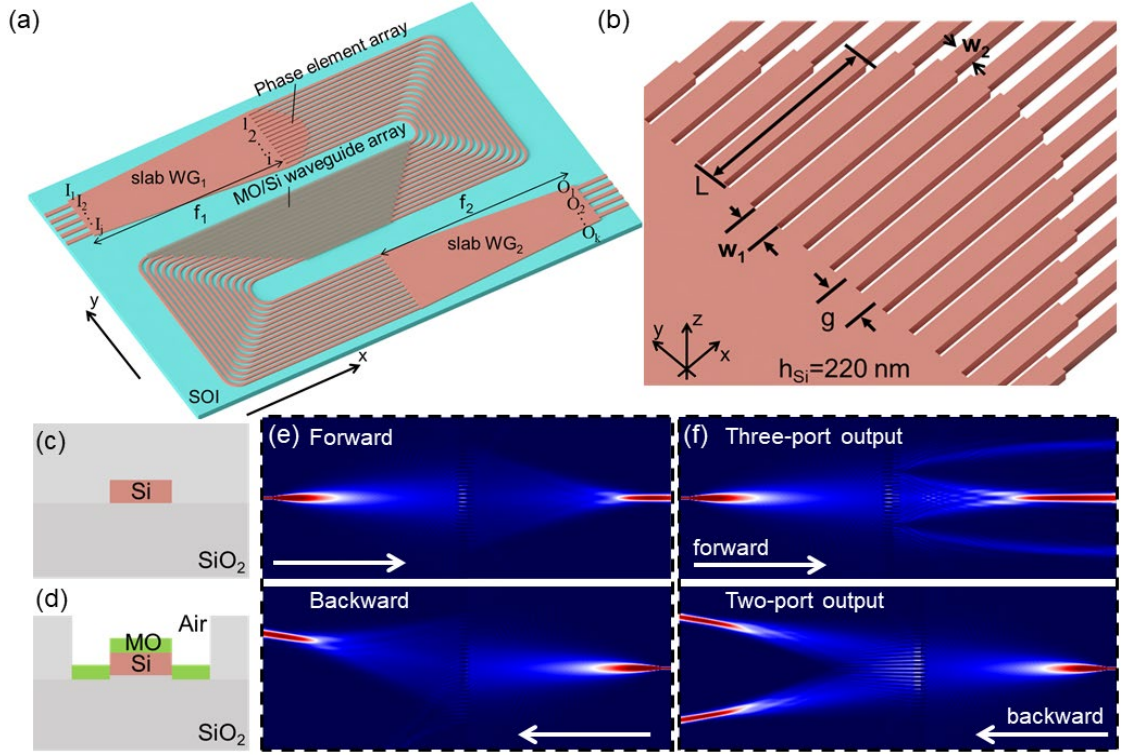

Fig.1 (a) Schematic of the proposed silicon integrated multi-port nonreciprocal optical device. (b) A sketch of the phase element array in the device. (c) Cross sectional structure of Si waveguides. (d) Cross sectional structure of MO/Si waveguides. (e) Simulated transmission power of a proposed optical circulator. (f) Simulated transmission power of a proposed multi-port MO device.

Let's consider the accumulated phase shift when light propagates from the input port $I_j$ to the output port $O_k$ through the $i_{th}$ waveguide in the waveguide array, as shown in Fig. 1(a). The total phase shift can be written as:

$$\varphi_i(fwd) = \varphi_{S(j,i)} + \varphi_{SOI(i)} + \varphi_{MO,i}^+ + \varphi_{S(k,i)} \tag{1}$$

Where $\varphi_{S(j,i)}$ is the phase shift when propagating from input port $I_j$ to the $i_{th}$ waveguide in the slab waveguide $WG_1$, $\varphi_{SOI(i)}$ is the reciprocal phase shift when propagating through the $i_{th}$ silicon waveguide, $\varphi_{MO,i}^+$ is the nonreciprocal phase shift when propagating through the $i_{th}$ MO/Si waveguide and $\varphi_{S(k,i)}$ is the phase shift when propagating from the $i_{th}$ waveguide to the output port $O_k$ in the slab waveguide $WG_2$. Constructive interference and high transmittance will be observed at the $O_k$ port when $\varphi_i$ between adjacent waveguides satisfies $2m\pi$ (m=0,1,2…) phase difference at the designed wavelength [22]. For backward propagation, due to nonreciprocal nature of the MO/Si waveguide the phase shift is:

$$\varphi_i(bwd) = \varphi_{S(j,i)} + \varphi_{SOI(i)} + \varphi_{MO,i}^- + \varphi_{S(k,i)} \tag{2}$$

Where $\varphi_{MO,i}^-$ is backward propagation phase shift of the MO waveguide, which differs from $\varphi_{MO,i}^+$ due to the nonreciprocal phase shift. In this case, different phase profile can be obtained by designing different lengths of the MO/Si waveguide array. Therefore, on-demand nonreciprocal transmission between forward and backward propagation can be achieved.

We can further express each of the phase terms as follows. $\varphi_{S(j,i)}$ and $\varphi_{S(k,i)}$ can be expressed as:

$$\varphi_{S(j,i)} = \frac{\omega}{c} n_{eff}^{Slab} \sqrt{(y_i - y_{I_j})^2 + f_1^2} \tag{3}$$

$$\varphi_{S(k,i)} = \frac{\omega}{c} n_{eff}^{Slab} \sqrt{(y_i - y_{O_k})^2 + f_2^2} \tag{4}$$

where ω is the frequency, c is the speed of light in vacuum, $n_{eff}^{Slab}$ is the effective index of the TM mode in the slab Si waveguides. ($y_i$-$y_{I_j}$) and ($y_i$-$y_{O_k}$) refers to the difference in the distance from the $j_{th}$ input port and $k_{th}$ output port to the $i_{th}$ arrayed waveguide along the y direction. $f_1$ and $f_2$ are the length of the input and output slab waveguides. Fig. 1(b) shows the zoom-in sketch of the phase element array, which is used to provide required reciprocal phase shift difference between waveguides in the array. The phase element consists of a wider waveguide ($w_1$) with length $L_{1,i}$ and effective index $n_{eff1}$ for the fundamental TM mode. The rest part of the silicon waveguides consists of a narrower waveguide ($w_2$) with length $L_{2,i}$ and effective index $n_{eff2}$ for the fundamental TM mode. The transmission phase of the $i_{th}$ silicon waveguide $\varphi_{SOI(i)}$ and nonreciprocal phase shift $\Delta\beta$ can be expressed as

$$\varphi_{SOI(i)} = \frac{\omega}{c} n_{eff1} L_{1,i} + \frac{\omega}{c} n_{eff2} L_{2,i} \tag{5}$$

$$\Delta\beta = \frac{2\beta}{\omega\varepsilon_0 N} \iint \frac{\gamma}{n_0^4} H_y \partial_z H_y dxdy \tag{6}$$

Fig. 1(d) shows the cross section of the MO/Si waveguide. The silicon dioxide ($SiO_2$) upper cladding in the MO/Si waveguide region was etched to expose the upper surface of the Si waveguide. This structure allowed CeYIG/YIG thin films to be directly deposited on top of Si waveguides, providing NRPS for the TM polarized light [26]. The transmission phase of MO/Si waveguides $\varphi_{MO,i}^{\pm}$ can be expressed as

$$\varphi_{MO,i}^{\pm} = \pm \Delta\beta L_{MO,i} + \frac{\omega}{c} n_{eff(MO)} L_{MO,i} \tag{7}$$

where $n_{eff}(MO)$ is the $TM_0$ mode effective index of the MO/Si waveguide under 0 applied magnetic field. Δβ and β are the NRPS and propagation constant of the fundamental TM mode in the MO/Si waveguide, respectively. ω is the angular frequency, γ is the off diagonal component of the permittivity tensor of the MO material. $\varepsilon_0$ is the vacuum dielectric constant, $N$ is the power flux along x direction, $n_0$ is the refractive index of the magneto-optical material, $H_y$ is the magnetic field along y direction. $L_{MO,i}$ is the length of the $i_{th}$ MO/Si waveguide. The length of the MO/Si waveguide $L_{MO,i}$ can be calculated by considering forward and backward transmission phase difference $\varphi_{MO,i}^{+} - \varphi_{MO,i}^{-}$ allowing forward and backward propagation light to be focused at different ports. The length of each MO/Si waveguide can be expressed as

$$L_{MO,i} = \frac{\varphi_{MO,i}^{+} - \varphi_{MO,i}^{-}}{\Delta\beta} \tag{8}$$

We simulated the transmission field profile of the device based on the above equations, as shown in Fig. 1(e) and 1(f). On-demand nonreciprocal optical routing can be achieved between different ports. For instance, an optical circulator can be realized as shown in Fig. 1(e). In this case, a phase gradient is introduced by the NRPS in the MO/Si waveguides for the backward incidence. The device operates as an optical circulator between the designed ports. It is possible to design different phase profiles of the waveguides so that more than one focusing point for forward and backward directions can be achieved. In this case, we can obtain multiport nonreciprocal optical routing as shown in Fig. 1(f). The forward transmission light is focused to three-ports, whereas the backward propagating light from one of the forward output ports is focused to two different ports. Therefore, nonreciprocal routing is achieved.

## 3. Device Design

The device was designed at 1550 nm wavelength. We first designed the slab waveguide regions, as shown in the simplified device structure in Fig. 2(a). The slab waveguide length $f_1$, $f_2$, the number of waveguides and the port spacing D influence the insertion loss and isolation ratio of the device. The forward and backward theoretical transmission phase gradients for the structure analysis is depicted in Fig. 2(b), corresponding to the case of Fig. 1(e) when the focal length $f_1$ and $f_2$ are initially set to be 200 μm. The forward (backward) transmission phase gradients indicate the phase distribution before focusing in the slab waveguide $WG_2$ ($WG_1$), consisting of the transmission phase accumulated in the $WG_1$ ($WG_2$) and the arrayed waveguides.

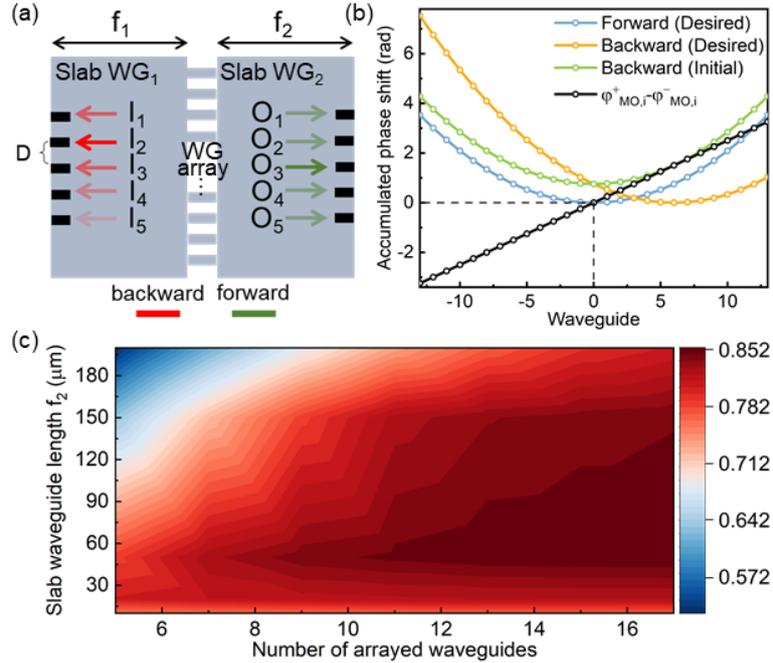

Fig.2 (a) A simplified schematic diagram of the proposed device. (b) The forward and backward theoretical transmission phase gradients of the proposed optical circulator using the focusing structure. (c) Forward transmission focusing efficiency changing with the number of the arrayed waveguides when the length of the output slab Si waveguide $f_2$ takes different values.

The focusing design requires that each of the arrayed waveguide transmission paths shares the same phase shift from the input to the output [22], which determines the desired forward phase gradients. Before the employment of NRPS, the phase gradients of the device in the backward transmission direction, which is shown as initial backward phase gradients in Fig. 2(b), can be derived utilizing the Lorentz reciprocity theorem [27]. The relative distribution of the initial backward phase gradients is the same as the forward one, while it is increased uniformly in terms of the absolute value for the following uniform NRPS distribution design (NRPS=0 in the center waveguide), which does not affect the light focusing and transmission states. Then the desired backward phase gradients for output focusing, which is calculated using Equations (3) and (4), can be realized by introducing NRPS to the initial backward phase gradients. The introduced NRPS $\varphi_{MO,i}^{+} - \varphi_{MO,i}^{-}$ in each arrayed waveguide is also shown in Fig. 2(b), by computing the disparity between the desired backward transmission phase gradients and the initial backward transmission phase gradients. The length of each MO/Si waveguide can be calculated accordingly using Equation (8). As shown in Fig. 2(b), the closer to the edge of the waveguide array, the larger the required NRPS. The largest required NRPS in the waveguide array is 3.26 rad, corresponding to the MO/Si waveguide length of 745.1 μm. Fig. 2(c) shows the forward focusing efficiency, defined as the transmission efficiency of light travelling from the slab waveguide $WG_2$ to the target output port, as a function of the number of the arrayed waveguides N and the slab waveguide length $f_2$. With increasing N from 6 to 17, the focusing efficiency monotonously increases from 60% to 85%. The focusing efficiency also changes with the length of the output slab Si waveguide $f_2$. A larger $f_2$ requires more waveguides to cover the phase gradient range of 2π. While a smaller $f_2$ leads to a sparser sampling of the phases with fixed waveguide spacing. The optimal $f_2$ is in the range of 50 μm to 100 μm, leading to the focusing efficiency of 85%. We fixed $f_2$=50 μm in our device design.

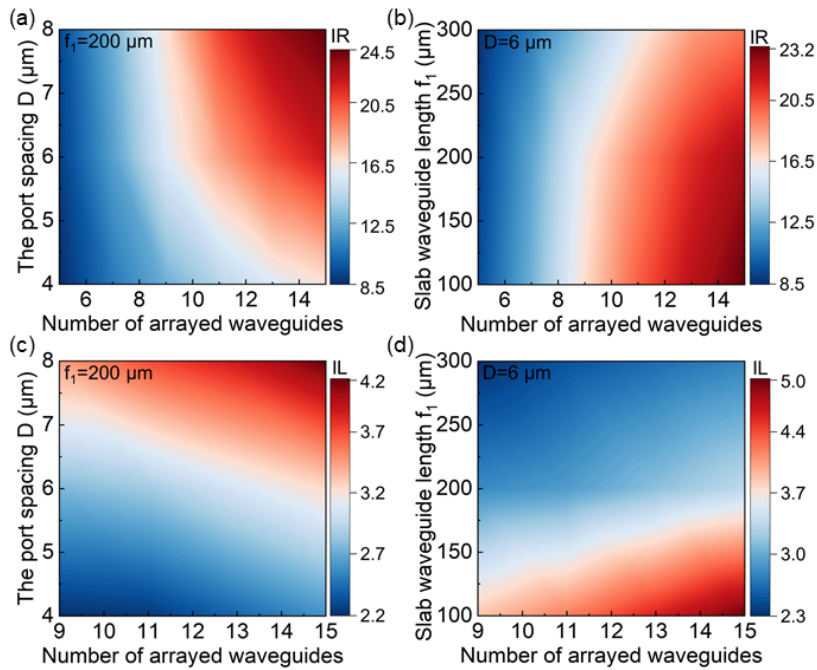

Fig.3 The (a) isolation ratio and the (c) insertion loss of the designed multi-port nonreciprocal optical device under different distance D between the backward output ports and number of arrayed waveguides when the length of the input slab Si waveguide $f_1$=200 μm. The (b) isolation ratio and the (d) insertion loss of the designed multi-port nonreciprocal optical device under different length of the input slab Si waveguide $f_1$ and number of arrayed waveguides when the distance between the backward output ports D=6 μm.

For backward propagation, the waveguide spacing D and Si slab waveguide length $f_1$ influence the isolation ratio and insertion loss. The isolation ratio is defined as the maximum transmission power difference between the two ports in the forward and backward transmission directions. Fig. 3(a) and (c) show simulated isolation ratio and insertion loss as a function of D and number of arrayed waveguides N for the case of Fig. 1(e). When the length of the input Si slab waveguide $f_1$ is fixed at 200 μm, the isolation ratio increases with the number of arrayed waveguides and spacing D between the backward output ports, as plotted in Fig. 3(a). As the focusing efficiency increases with the number of waveguides, and larger waveguide spacing D leads to weaker crosstalk from adjacent waveguides. However, larger D requires larger nonreciprocal phase gradient to allow efficient backward light focusing, which leads to longer MO/Si waveguides and larger insertion loss, as shown in Fig. 3(c). The theoretical insertion loss was obtained by calculating the focusing efficiency and total transmission loss in each arrayed waveguide including the Si and MO/Si waveguides. Under the same D=6 μm as shown in Fig. 3(b) and (d), smaller $f_1$ also requires a larger transmission phase difference between the forward and backward transmission directions, which results in a sharp increase in MO/Si waveguide loss. While at the same time, the limited number of arrayed waveguides could provide a larger phase gradient range for smaller $f_1$ in the backward transmission direction, which improves the isolation ratio to a certain extent. Considering the trade-off between the isolation ratio and insertion loss, we chose D=6 μm, $f_1$=200 μm, N=13, which leads to an isolation ratio of 21 dB and insertion loss of 3 dB. The insertion loss includes the focusing loss, junction loss and waveguide loss, where the waveguide loss consists of the transmission loss in the Si and MO/Si waveguides.

Next, we designed the waveguide arrays. The lengths of Si and MO/Si waveguides are determined by the forward and backward propagation phase profiles. For example, the case of Fig. 1(e) requires forward and backward phase profiles as shown in Fig. 2(b), including the transmission phase accumulated in the input slab waveguide and the arrayed waveguides. For the Si phase element arrays, we fixed the waveguide width at 1 μm and 500 nm., The 1 μm wide Si waveguide introduces reciprocal phase shift of 8.538 rad/μm. For MO/Si waveguides, the waveguide width is fixed at 500 nm, which introduces nonreciprocal phase shift Δβ of 4.375 rad/mm. In this case, the length of each arrayed phase element was fixed to be 6 μm, consisting of 500 nm and 1 μm Si waveguide sections. The length of the 1 μm wide Si waveguide can be adjusted from 0 to 6 μm. Phase shifts ranging from 0 to 2π can be achieved by changing the length ratio of wide and narrow waveguides in the phase element. The waveguide gap was designed at 1.5 μm to avoid strong mode coupling between Si waveguides. After passing through the 6-μm-long phase elements, the spacing between the Si waveguides was rapidly expanded to 5 μm to eliminate inter-waveguide crosstalk. The MO/Si waveguide array consists of MO/Si waveguides with different length, providing NRPS for the TM polarized light with MO thin films deposited on the top of Si channel waveguides.

Fig. 4 shows the illustration of nonreciprocal optical routing based on a 5×5 nonreciprocal router. In Fig. 4(a), the device is designed to allow the forward transmission as $I_1$ to $O_2$, whereas the backward transmission as $O_2$ to $I_3$, i.e., similar to a three port circulator. Fig. 4(b) shows the simulated transmission spectra during the process of light beam focusing. Theoretical isolation ratio of 21 dB and crosstalk of -21 dB were observed at 1550 nm wavelength, with total insertion loss of 3 dB, among which the focusing loss was 0.7 dB as shown in the simulation results in Fig. 4(b). The theoretical isolation ratio was mainly limited by the small distance D between the backward output ports. The backward focusing loss was slightly larger than the forward one because the phase gradient is no longer symmetrical compared to forward propagation, as shown in Fig. 3(b). When the wavelength deviated from the center 1550 nm wavelength, changes of the propagation constant in the Si and MO/Si waveguides led to the offset of the phase gradient, which resulted in the deterioration of the focusing efficiency and the isolation ratio.

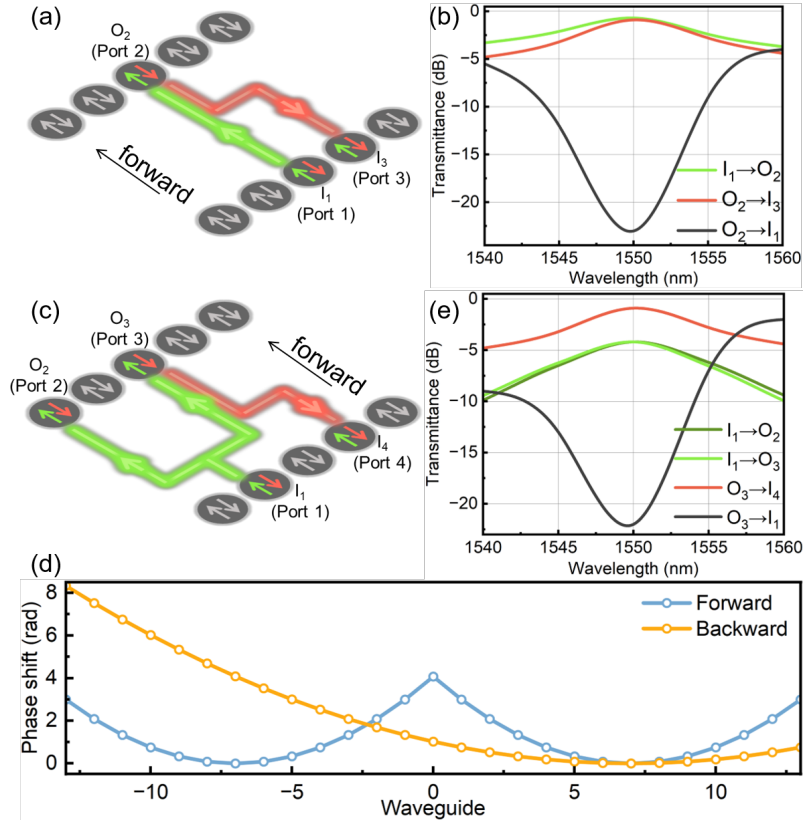

Fig.4 (a) Illustration of three-port optical circulation function based on the nonreciprocal optical focusing. (b) Theoretical transmission spectra of the three-port optical circulator. (c) Illustration of four-port nonreciprocal optical transmission. (d) Theoretical transmission spectra of the four-port nonreciprocal optical device. (e) The corresponding forward and backward theoretical transmission phase gradients of the proposed four-port device.

More complex nonreciprocal optical routing can be realized by dividing the waveguide array into different regions to focus the incident light to different ports. Fig. 4(c) shows an example. Light entering from $I_1$ is transmitted through both $O_2$ and $O_3$ for the forward, whereas the backward incident light from $O_3$ exits at $I_4$. This is realized by splitting the arrayed waveguides into two parts for two-port focusing in the forward transmission direction. Whereas for backward propagation, the phase gradients are changed through NRPS for another port focusing, as shown in Fig. 4(d). The phase gradients in the forward transmission direction exhibit a symmetrical design aimed at focusing the transmission light to two distinct outputs, separately. The backward phase gradients are tailored for single-port focusing, mirroring the configuration depicted in Fig. 2(b). The largest required NRPS in the waveguide array reaches 3.54 rad, corresponding to the MO/Si waveguide length of 810.9 μm. Theoretical transmission spectra of the device were shown in Fig. 4(e). We observed theoretical isolation ratio of 18 dB between $I_1$ and $O_3$, crosstalk of -21 dB between $I_1$ and $O_4$ at 1550 nm wavelength, with focusing loss of 0.9 dB and 1.2 dB for light transmission from $O_3$ to $I_4$ and from $I_1$ to $O_2/O_3$, respectively. The relatively higher focusing loss for forward transmission ($I_1 \rightarrow O_2/O_3$) was attributed to the reduction in the number of arrayed waveguides corresponding to each focusing point after the arrayed waveguides splitting. The theoretical insertion loss of the device could be then obtained to be 3.5 dB.

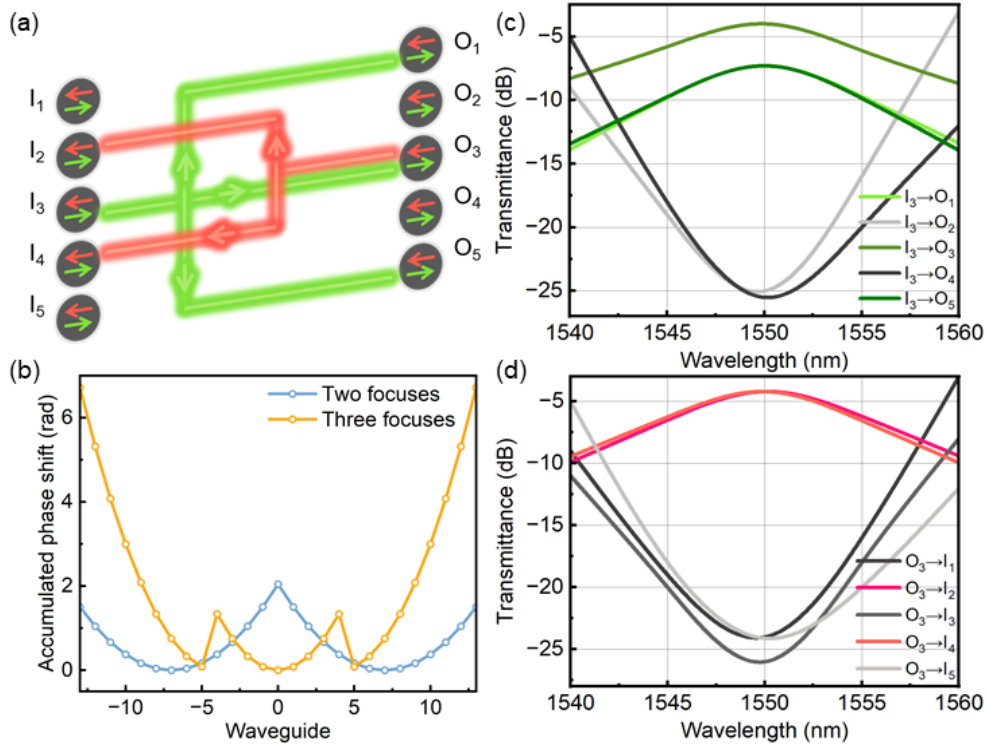

Fig.5 (a) Illustration of nonreciprocal optical transmission in a 5×5 silicon integrated multi-port nonreciprocal optical device. (b) The forward and backward theoretical transmission phase gradients of the proposed 5×5 multi-port nonreciprocal optical device. (c) Theoretical transmission spectra of the multi-port nonreciprocal optical device when the forward light was incident from $I_3$. (d) Theoretical transmission spectra of the multi-port nonreciprocal optical device when the backward light was incident from $O_3$.

More complicate nonreciprocal transmission can be realized as shown in Fig. 5(a). When the forward propagating light is incident from $I_3$, it transmitted to $O_1$, $O_3$, and $O_5$, respectively. While the light is backward incident from $O_3$, it transmitted to $I_2$ and $I_4$. The phase gradients of the waveguide array for the forward and backward transmission directions for the case of Fig. 1(f) is shown in Fig. 5(b). In the corresponding transmission direction, the arrayed waveguides are divided into several parts according to the number of target ports to be focused. Then the forward and backward transmission phase in each part of the arrayed waveguides can be separately calculated as shown in Fig. 5(b). Theoretical transmission spectra of the device with the incident light from $I_3$ and $O_3$ are shown in Fig. 5(c) and (d). A minimum crosstalk of -21 dB between $O_3$ and $O_4$ could be achieved. The theoretical isolation ratio reached 21 dB between $I_3$ and $O_3$ at 1550 nm wavelength. The focusing loss from $I_3$ to $O_3$ was about 1.5 dB because only half of the arrayed waveguides were designed for the light transmitting to $O_3$. The focusing loss from $O_3$ to $I_2/I_4$ was about 1.2 dB, which was similar to the situation shown in Fig. 4(d). Considering the corresponding focusing loss, the theoretical insertion loss of the device was 3.8 dB.

## 4. Device Fabrication and Characterization

The proposed devices were fabricated in a silicon photonics foundry followed by MO thin film deposition. The Si waveguide was prepared on a 220 nm silicon on insulator (SOI) wafer by reactive ion etching (RIE). The $SiO_2$ top cladding was then etched to expose the silicon waveguide core. Yttrium iron garnet (YIG) and cerium substituted yttrium iron garnet (Ce:YIG) thin films were deposited to form the MO/Si waveguides. Details of the fabrication process can be referred to our previous publications [8]. The transmission spectra were then measured under a 1000 Oe in-plane applied magnetic field on a polarization maintaining, fiber-butt-coupled system as detailed in [8].

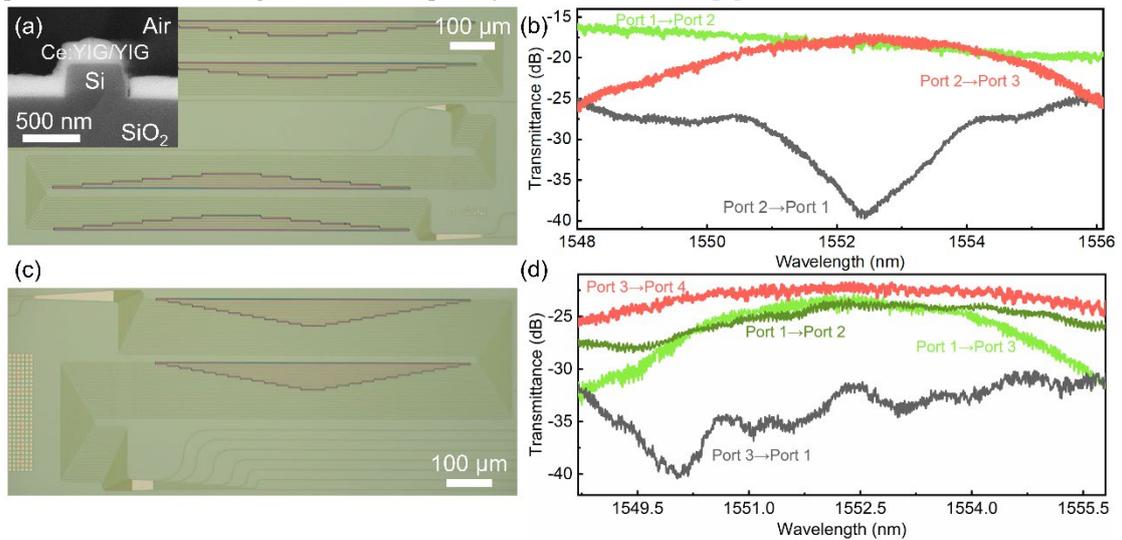

Fig.6 (a) Optical microscope image of the fabricated three-port optical circulator. The inset is the scanning electron microscope (SEM) image of the cross-sectional structure of MO/Si waveguide. (b) Experimentally measured transmission spectra of the three-port optical circulator. (c) Optical microscope image of the fabricated four-port nonreciprocal optical device. (d) Experimentally measured transmission spectra of the four-port nonreciprocal optical device.

Optical microscope image of the fabricated three-port optical circulator is shown in Fig. 6(a). The inset shows the scanning electron microscope (SEM) image of the cross-sectional structure of MO/Si waveguide, which is consistent with Fig. 1(d). Fig. 6(b) shows the measured transmission spectra of the three-port optical circulator. The device achieved 19 dB isolation ratio between port 1 and port 2 with -19 dB crosstalk between port 1 and port 3 at 1552.3 nm wavelength. The device insertion loss was 12 dB, which was mainly attributed to the junction loss between the Si and MO/Si waveguide (3.5 dB), Si and MO/Si waveguide loss (2.5 dB), and focusing loss (6 dB) due to the fabrication caused non-ideal phase gradients. Fig. 6(c) shows the optical microscope image of the fabricated four-port nonreciprocal optical router for the case of Fig. 4(c). The corresponding measured transmission spectra of the device is shown in Fig. 6(d). The forward optical transmission power from port 1 to ports 2 and 3 was equally divided at 1552.7 nm. At 1549.9nm wavelength, the device showed a maximum isolation ratio of 12 dB between port 1 and port 3 with a minimum crosstalk of about -16 dB between port 4 and port 1. The minimum insertion loss of the device at 1552.5 nm was 15 dB, mainly attributed to the junction loss between the Si and MO/Si waveguides (3.5 dB), Si and MO/Si waveguide loss (3 dB), and focusing loss (8.5 dB) due to the fabrication caused non-ideal phase gradients. The insertion loss of the device could be further optimized through the taper designs at the junctions between slab Si waveguides and arrayed Si waveguides as well as precise phase control based on the thermo-optic effects.

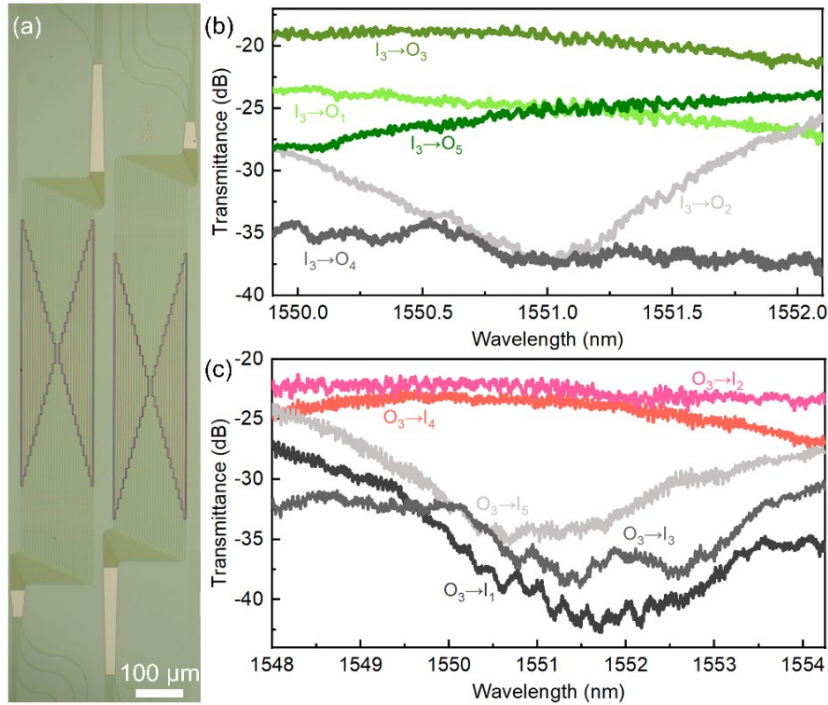

Fig.7 (a) Optical microscope image of the fabricated 5×5 multi-port nonreciprocal optical device. (b) Experimental measured transmission spectra of the 5×5 multi-port nonreciprocal optical device when the forward light was incident from $I_3$. (c) Experimental measured transmission spectra of the 5×5 multi-port nonreciprocal optical device when the backward light was incident from $O_3$.

Finally, the optical microscope image of a 5×5 nonreciprocal optical router for the case of Fig. 1(f) is shown in Fig. 7(a). The measured transmission spectra of the device are shown in Fig. 7(b) and 7(c). An isolation ratio up to 16 dB between $I_3$ and $O_3$ with a minimum crosstalk of -18 dB between $I_1$ and $I_2$ was achieved at 1551 nm wavelength.

The high insertion loss of fabricated devices mainly includes the junction loss, MO/Si waveguide loss and the focusing loss due to phase error. The junction loss and MO/Si waveguide loss can be further minimized by employing tapered structures and optimized MO thin films [14]. While the phase offset due to fabrication error also seriously affects the focusing efficiency in experiments because of the long transmission length of the phased waveguide array in the device. When the fabricated width of the silicon waveguide changes 5% from the designed value of 500 nm to 475 nm in the case of Fig. 6(a), the phase shift offset due to the fabrication error reached nearly 12 rad. Compared with the theoretical ideal focusing efficiency of 85%, the focusing efficiency considering the fabrication error drops to about 25%. The high focusing loss can be mitigated by shortening the silicon waveguide length with push-pull magnetic field application [10] and using thermo-optic heating to compensate the phase errors [26].

## 5. Conclusion

In summary, we experimentally demonstrated silicon integrated multi-port nonreciprocal optical devices at near infrared optical communication wavelengths. On-desire multi-port nonreciprocal optical routing has been demonstrated in a 5×5 magneto-optical nonreciprocal router integrated on SOI waveguides. The fabricated device showed -18 dB crosstalk and 16 dB isolation ratio at 1551 nm wavelength. Such multi-port nonreciprocal optical devices demonstrated in this work allow different scattering matrix for forward and backward incidence, adding a new degree of freedom for optical communication and data communication applications.

**Funding.**

The authors are grateful for the support by the Ministry of Science and Technology of the People's Republic of China (MOST) (Grant No. 2021YFB2801600), National Natural Science Foundation of China (NSFC) (Grant Nos. U22A20148, 51972044, 52021001 and 52102357), Sichuan Provincial Science and Technology Department (Grant Nos. 99203070, 23ZYZYTS0043).

**Author contributions.**

X. S., W. Y. and D. W. contributed equally to this work by jointly searching the data, performing simulations and experiments, and writing the manuscript. Y. Y., Z. W., Z. Z., T. Z., and J. W. contributed to the discussion. J. Q. and L. B. conceived the study and supervised the project. All the authors contributed to the final version of this manuscript.

## Disclosures.

The authors declare no conflicts of interest.

## Reference

1. Wang, X., H. Yu, H. Qiu, Q. Zhang, Z. Fu, P. Xia, B. Chen, X. Guo, Y. Wang, X. Jiang, and J. Yang, "Hitless and gridless reconfigurable optical add drop (de)multiplexer based on looped waveguide sidewall Bragg gratings on silicon," Optics Express **28**, 14461-14475 (2020).
2. Liu, D., L. Zhang, H. Jiang, and D. Dai, "First demonstration of an on-chip quadplexer for passive optical network systems," Photonics Research **9**, 757-763 (2021).
3. Li, C., D. Liu, and D. Dai, "Multimode silicon photonics," Nanophotonics **8**, 227-247 (2019).
4. Park, J., M.-J. Kwack, J. Joo, and G. Kim, "Low-loss single-mode operation in silicon multi-mode arrayed waveguide grating with a double-etched inverse taper structure," Journal of Optics **19**, (2017).
5. Chen, S., Y. Shi, S. He, and D. Dai, "Compact Eight-Channel Thermally Reconfigurable Optical Add/Drop Multiplexers on Silicon," IEEE Photonics Technology Letters **28**, 1874-1877 (2016).
6. Yang, K.Y., J. Skarda, M. Cotrufo, A. Dutt, G.H. Ahn, M. Sawaby, D. Vercruysse, A. Arbabian, S. Fan, A. Alù, and J. Vučković, "Inverse-designed non-reciprocal pulse router for chip-based LiDAR," Nature Photonics **14**, 369-374 (2020).
7. Liu, S., D. Minemura, and Y. Shoji, "Silicon-based integrated polarization-independent magneto-optical isolator," Optica **10**, (2023).
8. Yan, W., Y. Yang, S. Liu, Y. Zhang, S. Xia, T. Kang, W. Yang, J. Qin, L. Deng, and L. Bi, "Waveguide-integrated high-performance magneto-optical isolators and circulators on silicon nitride platforms," Optica **7**, (2020).
9. Shoji, Y., A. Fujie, and T. Mizumoto, "Silicon Waveguide Optical Isolator Operating for TE Mode Input Light," IEEE Journal of Selected Topics in Quantum Electronics **22**, 264-270 (2016).
10. Shoji, Y. and T. Mizumoto, "Silicon Waveguide Optical Isolator with Directly Bonded Magneto-Optical Garnet," Applied Sciences **9**, (2019).
11. Yamaguchi, R., Y. Shoji, and T. Mizumoto, "Low-loss waveguide optical isolator with tapered mode converter and magneto-optical phase shifter for TE mode input," Optics Express **26**, 21271-21278 (2018).
12. Liu, S., Y. Shoji, and T. Mizumoto, "Mode-evolution-based TE mode magneto-optical isolator using asymmetric adiabatic tapered waveguides," Optics Express **29**, 22838-22846 (2021).
13. Huang, D., P. Pintus, C. Zhang, P. Morton, Y. Shoji, T. Mizumoto, and J.E. Bowers, "Dynamically reconfigurable integrated optical circulators," Optica **4**, 23-30 (2016).
14. Du, Q., C. Wang, Y. Zhang, Y. Zhang, T. Fakhrul, W. Zhang, C. Gonçalves, C. Blanco, K. Richardson, L. Deng, C.A. Ross, L. Bi, and J. Hu, "Monolithic On-chip Magneto-optical Isolator with 3 dB Insertion Loss and 40 dB Isolation Ratio," ACS Photonics **5**, 5010-5016 (2018).
15. Liu, S., Y. Shoji, and T. Mizumoto, "TE-mode magneto-optical isolator based on an asymmetric microring resonator under a unidirectional magnetic field," Optics Express **30**, 9934-9943 (2022).
16. Lu, Y., K. Xu, Z. Lu, H. Yang, Y. Zhai, and M. Bi, "Bidirectional passive optical interconnected

data center," Microwave and Optical Technology Letters **66**, e34133 (2024).
17. Tsalamanis, I., E. Rochat, and S.D. Walker, "Experimental demonstration of cascaded AWG access network featuring bi-directional transmission and polarization multiplexing," Optics Express **12**, 764-769 (2004).
18. Liu, D., L. Zhang, H. Jiang, and D. Dai, "First demonstration of an on-chip quadplexer for passive optical network systems," Photonics Research **9**, (2021).
19. Martin, A., P. Verheyen, P. De Heyn, P. Absil, P. Feneyrou, J. Bourderionnet, D. Dodane, L. Leviandier, D. Dolfi, A. Naughton, P. O'Brien, T. Spuessens, R. Baets, and G. Lepage, "Photonic Integrated Circuit-Based FMCW Coherent LiDAR," Journal of Lightwave Technology **36**, 4640-4645 (2018).
20. Zhou, H., J. Chee, J. Song, and G. Lo, "Analytical calculation of nonreciprocal phase shifts and comparison analysis of enhanced magneto-optical waveguides on SOI platform," Optics Express **20**, 8256-8269 (2012).
21. Zhang, J., J. Yang, H. Xin, J. Huang, D. Chen, and Z. Zhaojian, "Ultrashort and efficient adiabatic waveguide taper based on thin flat focusing lenses," Optics Express **25**, 19894-19903 (2017).
22. Wang, Z., T. Li, A. Soman, D. Mao, T. Kananen, and T. Gu, "On-chip wavefront shaping with dielectric metasurface," Nature Communications **10**, 3547 (2019).
23. Liao, K., T. Gan, X. Hu, and Q. Gong, "AI-assisted on-chip nanophotonic convolver based on silicon metasurface," Nanophotonics **9**, 3315-3322 (2020).
24. Chen, W.T., A.Y. Zhu, V. Sanjeev, M. Khorasaninejad, Z. Shi, E. Lee, and F. Capasso, "A broadband achromatic metalens for focusing and imaging in the visible," Nat Nanotechnol **13**, 220-226 (2018).
25. Chen, W.T., A.Y. Zhu, J. Sisler, Z. Bharwani, and F. Capasso, "A broadband achromatic polarization-insensitive metalens consisting of anisotropic nanostructures," Nature Communications **10**, 355 (2019).
26. Yan, W., Z. Wei, Y. Yang, D. Wu, Z. Zhang, X. Song, J. Qin, and L. Bi, "Ultra-broadband magneto-optical isolators and circulators on a silicon nitride photonics platform," Optica **11**, 376-384 (2024).
27. Jalas, D., A. Petrov, M. Eich, W. Freude, S. Fan, Z. Yu, R. Baets, M. Popović, A. Melloni, J.D. Joannopoulos, M. Vanwolleghem, C.R. Doerr, and H. Renner, "What is — and what is not — an optical isolator," Nature Photonics **7**, 579-582 (2013).